\documentclass[a4paper,11pt]{article}
\usepackage{pos}
\usepackage{physics}

\newcommand{\bvec}[1]{\boldsymbol{#1}}

\title{Study of $SU(2)$ gauge theories with multiple Higgs fields in different representations}

\author[a]{Guilherme Catumba}
\author[b]{Atsuki Hiraguchi}
\author[c]{George W.-S. Hou}
\author[d]{Karl Jansen}
\author[c,e]{Ying-Jer Kao}
\author[b,f,g,h]{C.-J. David Lin}
\author[a]{Alberto Ramos}
\author*[c,h]{Mugdha Sarkar}

\affiliation[a]{Instituto de Física Corpuscular (IFIC), CSIC-Universitat de Valencia.\\
46071, Valencia, Spain}

\affiliation[b]{Institute of Physics, National Yang Ming Chiao Tung University,\\ 
1001 Ta-Hsueh Road, Hsinchu 30010, Taiwan}

\affiliation[c]{Department of Physics, National Taiwan University,\\
Taipei 10617, Taiwan}

\affiliation[d]{Deutsches Elektronen-Synchrotron DESY,\\
Platanenallee 6, 15738 Zeuthen, Germany}

\affiliation[e]{Center for Theoretical Physics and Center for Quantum Science and Technology, National Taiwan University,\\
Taipei, 10607, Taiwan}
  
\affiliation[f]{Center for High Energy Physics, Chung-Yuan Christian University,\\
Chung-Li 32023, Taiwan}

\affiliation[g]{Centre for Theoretical and Computational Physics, National Yang Ming Chiao Tung University,\\ 
1001 Ta-Hsueh Road, Hsinchu 30010, Taiwan}

\affiliation[h]{Physics Division, National Centre for Theoretical Sciences,\\ 
Taipei 10617, Taiwan}

\emailAdd{mugdha.sarkar@phys.ncts.ntu.edu.tw}

\abstract{We study two different SU(2) gauge-scalar theories in 3 and 4 spacetime
dimensions. Firstly, we focus on the 3 dimensional SU(2) theory with
multiple Higgs fields in the adjoint representation, that can be mapped to cuprate
systems in condensed matter physics which host a rich phase diagram including
high-Tc superconductivity. It has been proposed that the theory with 4 adjoint Higgs
fields can be used to explain the physics of hole-doped cuprates for a wide range
of parameters. We show exploratory results on the phase diagram of the theory.

On the other hand, we are interested in the 4 dimensional theory with 2 sets of
fundamental scalar (Higgs) fields, which is relevant to the 2 Higgs Doublet Model
(2HDM), a proposed extension to the Standard Model of particle physics.
The goal is to understand the particle spectrum of the theory at zero temperature
and the electroweak phase transition at finite temperature. We present exploratory
results on scale setting and the multi-parameter phase diagram of this theory.}

\FullConference{%
The 39th International Symposium on Lattice Field Theory,\\
8th-13th August, 2022,\\
Rheinische Friedrich-Wilhelms-Universität Bonn, Bonn, Germany
}

\begin{document}
\maketitle
\section{Introduction}
Lattice gauge theories coupled to scalar (Higgs) fields have been studied 
for several decades, due to their applicability in 
diverse fields of physics.
We study two $SU(2)$ gauge theories coupled to Higgs fields in different 
representations, one being relevant to BSM physics and the other to 
high-$T_c$ superconducting cuprate systems.
The lattice action for $D$-dimensional $SU(2)$ gauge theory with $N_h$ Higgs 
fields transforming in a given representation of the gauge group can be written as
\begin{gather}
S = \beta \sum_x \sum_{\mu < \nu}^D \left(1 - \frac{1}{2} \Tr U_{\mu\nu}(x) \right) 
   - \sum_x \sum_{\mu=1}^D \sum_{n=1}^{N_h} \sum_{i,j=1}^{N_g} \kappa_n (\Phi_n^i(x))^\ast \tilde{U}_\mu^{ij}(x)\Phi_n^j(x+\hat\mu)  \nonumber \\
   + \sum_x \sum_{n=1}^{N_h} \sum_{i=1}^{N_g} (\Phi_n^i(x))^\ast \Phi_n^i(x)
   + \sum_x \sum_{n=1}^{N_h} \lambda_n \left( \sum_{i=1}^{N_g}(\Phi_n^i(x))^\ast\Phi_n^i(x) - 1\right)^2 
   + V(\{\Phi_n\}) \, ,\label{Eq:genaction}
\end{gather}
where the first term is the usual Wilson plaquette action of gauge links $U_\mu(x)$ in 
the fundamental representation and $\tilde{U}_\mu^{ij}(x)$ indicates 
the $SU(2)$ gauge link in a given representation, with the indices 
$i,j=1,\ldots,N_g$, $N_g$ being the dimension of the said representation. 
The dimensionless Higgs fields $\Phi_n^i(x)$ are real/complex-valued 
depending on the representation, with $n=1\ldots N_h$ being 
the flavor index and $i=1\ldots N_g$ being the representation index. 
Depending on the global symmetries of the 
theory, there can be other quartic interaction terms for the Higgs fields which are 
denoted by $V(\{\Phi_n\})$.

\section{Gauge theory for cuprates on the lattice}
Since their discovery in the 1980s, cuprate superconductors have been a topic of 
tremendous interest 
in the condensed matter community. Along with high temperature superconductivity 
(at $\sim$ 150 $K$), these materials exhibit rich phases 
with the variation of temperature, doping densities, magnetic field, etc.
Of special interest are hole-doped cuprates near optimal doping, where 
there exist a plethora of 
unconventional phases with charge density wave (CDW) and/or nematic order that 
are unveiled from beneath the superconducting phase at strong magnetic field \cite{CMPreview}. 
It is 
imperative to understand the microscopic mechanism behind these phases to 
explain the superconducting phenomenon. 

In Ref.~\cite{Sachdev:2018nbk}, the authors proposed a $SU(2)$ gauge 
theory with $N_h=4$ Higgs fields transforming under the adjoint representation 
of the gauge group in ($2+1$) spacetime dimensions to explain the $2d$ 
hole-doped cuprate systems around optimal doping. The usual confining phase of 
the gauge theory corresponds to the Fermi liquid phase at 
high doping density while the broken (Higgs) phase can be mapped to the 
unconventional phases at underdoping including the pseudogap phase. 
In contrast to the $N_h=1$ case \cite{Hart:1996ac}, the $SU(2)$ gauge symmetry is expected to be 
broken into either $U(1)$-symmetric or $Z(2)$-symmetric Higgs phases.

The continuum action of the theory is given in Ref.~\cite{Sachdev:2018nbk}.
We aim to study the phase diagram by discretising the continuum action 
on a $3d$ Euclidean spacetime lattice. 
In Eq.~\ref{Eq:genaction}, we set $D=3$, $N_h=4$ and the 
couplings $\kappa_n=\kappa$ and $\lambda_n=\lambda$ for this theory. 
The $3\times 3$ adjoint representation link variable matrix 
can be written in terms of 
the $2\times 2$ fundamental link variable as 
$\tilde{U}_\mu^{ij}(x)=2\Tr (T^i U_\mu(x) T^j U_\mu^\dagger(x))$, 
where $T^i=\sigma_i/2$ are the generators of $SU(2)$ in the fundamental 
representation, $\sigma_i$ being the Pauli matrices. 
For convenience, we separate the complex Higgs 
fields $\mathcal{H}_{x/y}$ defined in \cite{Sachdev:2018nbk} as
  $\bvec{\mathcal{H}}_{x} = \varphi_1 + i\varphi_2$,
  $\bvec{\mathcal{H}}_{y} = \varphi_3 + i\varphi_4$,
to four real 3-component adjoint Higgs fields 
$\varphi_n^i(x)$ ($n=1,\ldots,4$ 
and $i=1,2,3$). The dimensionless Higgs fields for the lattice action 
(cf.~Eq.~\ref{Eq:genaction}) is related to the dimensionful $\varphi_n^i(x)$ 
as $\Phi_n^i = \sqrt{a/\kappa} \varphi_n^i$ , with $a$ being the lattice spacing.
Since there are only local terms in the Higgs potential, 
we suppress the spacetime index and write the potential as
  \begin{align}
  V(\{\Phi_n\}) = 4\lambda \sum_x&\sum_{n \neq m}^4 V_{mm}V_{nn} 
  + \hat{u_1}\sum_x \sum_{m,n}^4 g_{mn} V_{mm}V_{nn}
  + 4\hat{u_3}\sum_x \sum_{n \neq m}^4 V_{mn}V_{mn} \nonumber \\
  + \hat{u_2}\sum_x&  \left[\sum_{n=1}^4 (V_{nn})^2 
  + 4(V_{12})^2 + 4(V_{34})^2 - 2V_{11}V_{22}
  - 2V_{33}V_{44} \right]  \,, 
  \end{align}
where we have used the abbreviation $V_{mn}=\Tr(\Phi_m\Phi_n)$ and the 
Higgs fields have been expressed as $\Phi_n=\Phi_n^i T^i$ such that
$2\Tr(\Phi_m\Phi_n)=\sum_{i=1}^{3}\Phi_m^i\Phi_n^i$. In the 
2\textsuperscript{nd} term, the $g_{mn}$ are symmetric
and given by $g_{mm}=g_{12}=g_{34}=1$ and -1 for all other combinations. 
The dimensionful continuum couplings in Eq.~(2.1c) of
Ref.~\cite{Sachdev:2018nbk} are related 
to the dimensionless lattice couplings as follows
  \begin{gather} \label{Eq:cont_lat_relations_nh4}
  \beta = \frac{4}{g^2 a}, \quad \lambda=u_0 a \kappa^2, 
  \quad s = \frac{1}{a^2}\left[\frac{1}{\kappa} - 3 - 
  \frac{2\lambda}{\kappa}\right], \quad
  \hat{u_i} = u_i a \kappa^2 \quad (i=1,2,3).
  \end{gather}

\begin{figure}
  \centering
   \includegraphics[width=0.39\textwidth]{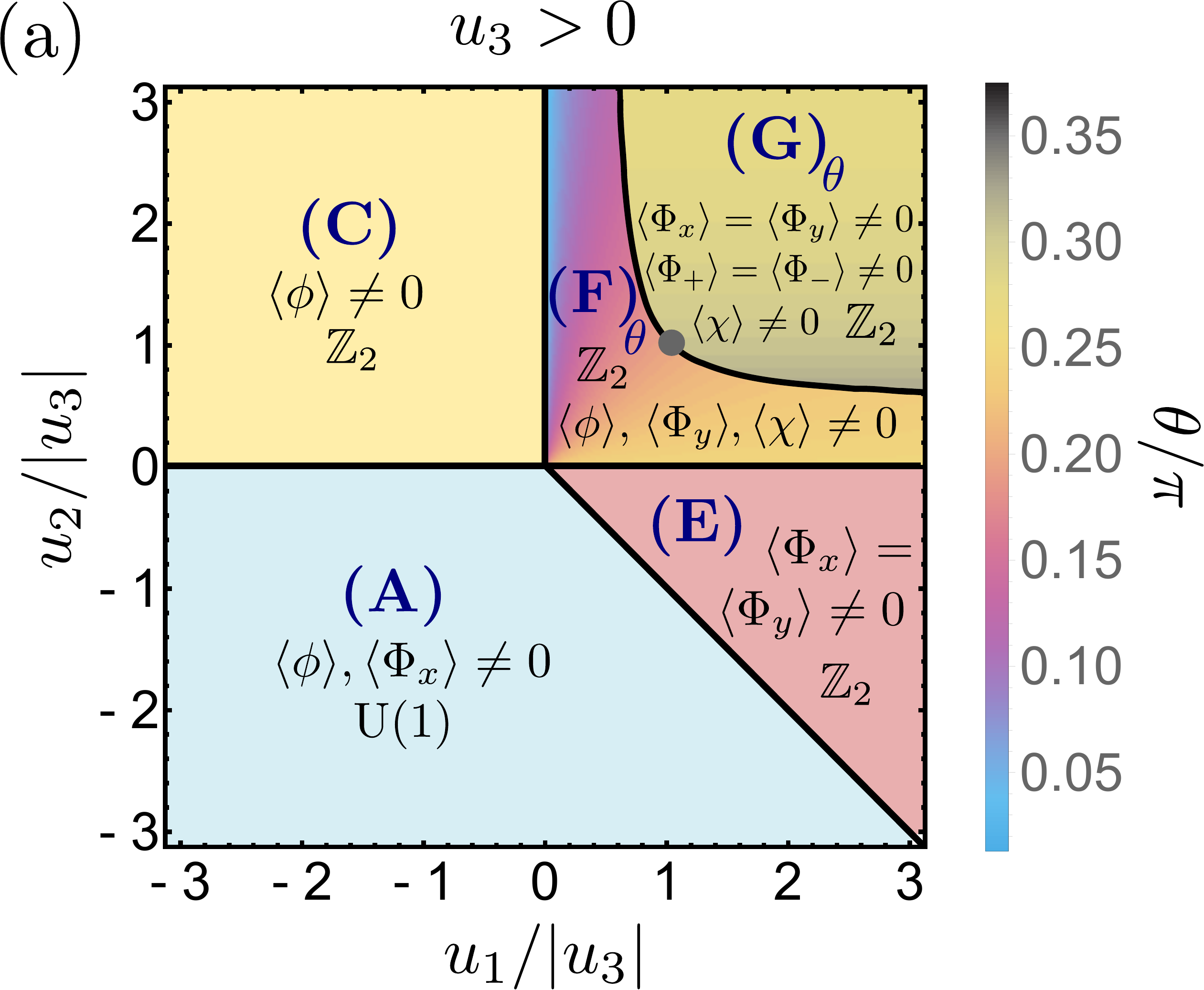}
   \includegraphics[width=0.2925\textwidth]{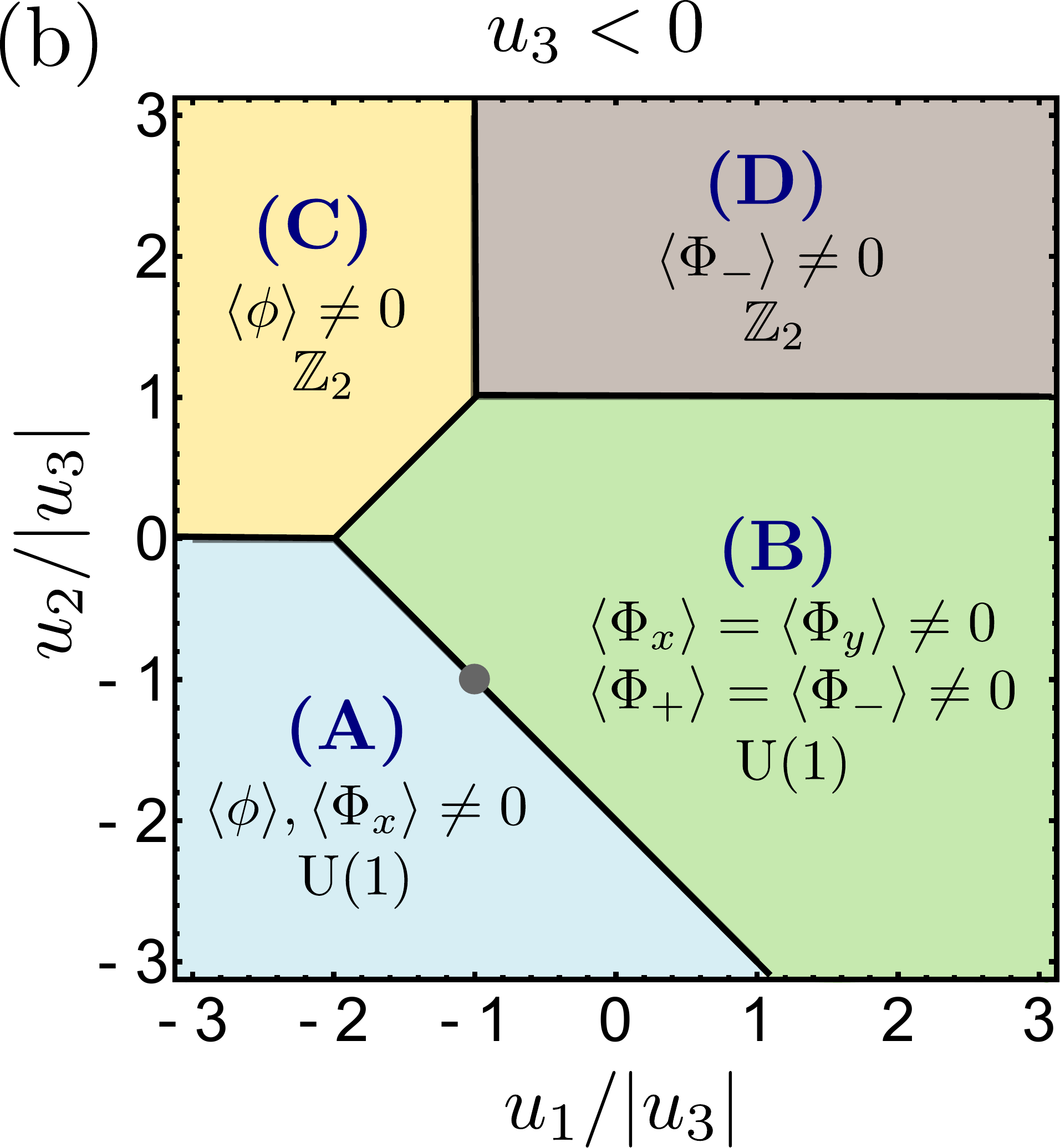}
	\caption{{Mean-field prediction of the various broken phases in the 
  3d $SU(2)$ gauge theory with $N_h=4$ adjoint Higgs,
  taken from Ref.~\cite{Sachdev:2018nbk}. The expectation 
  values of the observables for the different phases depicted above correspond to a 
  particular choice in which the adjoint Higgs fields get the vev and there are 
  other symmetry-equivalent choices. In general, the observables 
  $\Phi_{x,y}$ and $\Phi_{+,-}$ can be 
  complex and hence, we consider the modulus of them in our study. Moreover, the 
  phases with non-zero $\expval{\Phi_x}$ ($\expval{\Phi_+}$) can also  
  have a non-zero $\expval{\Phi_y}$ ($\expval{\Phi_-}$) for other choices of the vev.
  In plot (b), the expectation values of $\Phi_{x,y}$ and $\Phi_{+,-}$ in phase (B) 
	are all equal.}
	} \label{fig:meanfld}
 \end{figure}

The terms in the Higgs potential are constrained by the global symmetries 
of the hole-doped cuprate systems. The breaking of these symmetries leads to 
different phases within the Higgs phase. These Higgs phases,
labelled by (A)-(G) in Fig.~\ref{fig:meanfld}, can be 
characterised by various gauge-invariant bilinear combinations\footnote{{The observable 
$\chi$ in the left plot of Fig.~\ref{fig:meanfld} is a trilinear in the Higgs fields 
and is being currently investigated.}} 
combinations of the adjoint Higgs fields (cf.~\cite{Sachdev:2018nbk}),
\begin{align}
\phi = \Phi_1^i\Phi_1^i + \Phi_2^i\Phi_2^i - \Phi_3^i\Phi_3^i - \Phi_4^i\Phi_4^i  
\quad &\text{(Ising nematic order)},\nonumber \\
\Phi_x = \Phi_1^i\Phi_1^i - \Phi_2^i\Phi_2^i + 2i\Phi_1^i\Phi_2^i \quad &(\text{CDW order at wave vector } K_x), \nonumber \\
\Phi_y = \Phi_3^i\Phi_3^i - \Phi_4^i\Phi_4^i + 2i\Phi_3^i\Phi_4^i \quad &(\text{CDW order at wave vector } K_y), \nonumber \\
\Phi_+ = \Phi_1^i\Phi_3^i - \Phi_2^i\Phi_4^i + i(\Phi_1^i\Phi_4^i + \Phi_2^i\Phi_3^i)\quad &(\text{CDW order at wave vector } K_x+K_y), \nonumber \\
\Phi_- = \Phi_1^i\Phi_3^i + \Phi_2^i\Phi_4^i + i(\Phi_1^i\Phi_4^i - \Phi_2^i\Phi_3^i)\quad &(\text{CDW order at wave vector } K_x-K_y),
	\label{Eq:bilinear}
\end{align}
where the sum over the adjoint index $i$ is implied.  
The different broken phases in the Higgs phase have been 
predicted through mean-field analysis \cite{Sachdev:2018nbk} 
and is shown in Fig.~\ref{fig:meanfld}. In the analysis, 
the coupling $s$ is chosen negative enough to be in the 
broken phase and the quartic coupling $u_0$ is chosen large enough to 
stabilise the potential. A follow-up work \cite{Scammell:2019erm}
studied the model numerically in the strong gauge coupling limit with the 
global $O(N_h=4)$-invariant action obtained by setting the couplings 
$u_1=u_2=u_3$. In this approximation, the existence of the confinement phase 
and broken phases, symmetric under $U(1)$ and $Z(2)$,
were confirmed. Another recent study \cite{Bonati:2021tvg} in the 
$O(4)$ limit also demonstrated the existence of the two differently-broken 
Higs phases. In this work, we study the complete action, proposed in 
\cite{Sachdev:2018nbk}, non-perturbatively for the first time.

\begin{figure}
  \centering
   \includegraphics[width=0.37\textwidth]{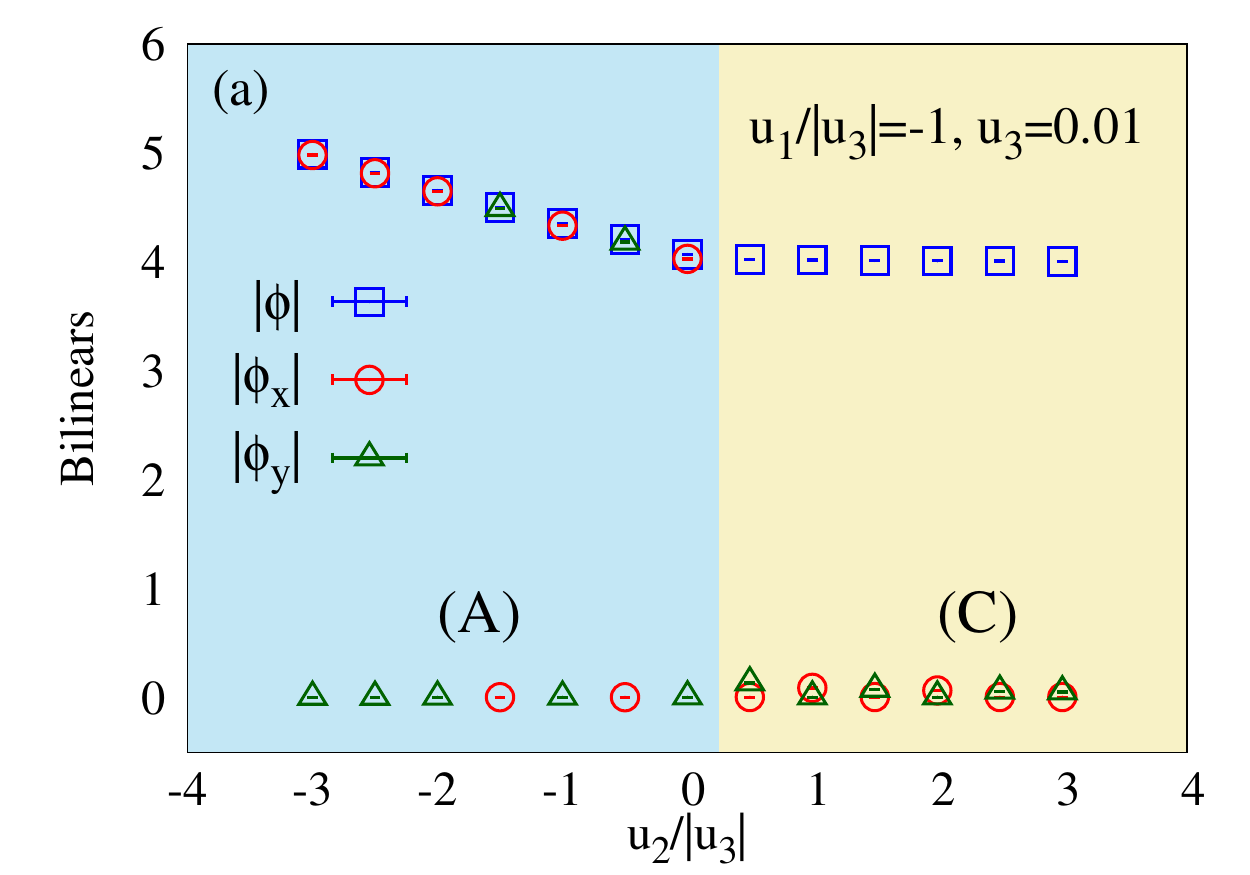}
   \includegraphics[width=0.37\textwidth]{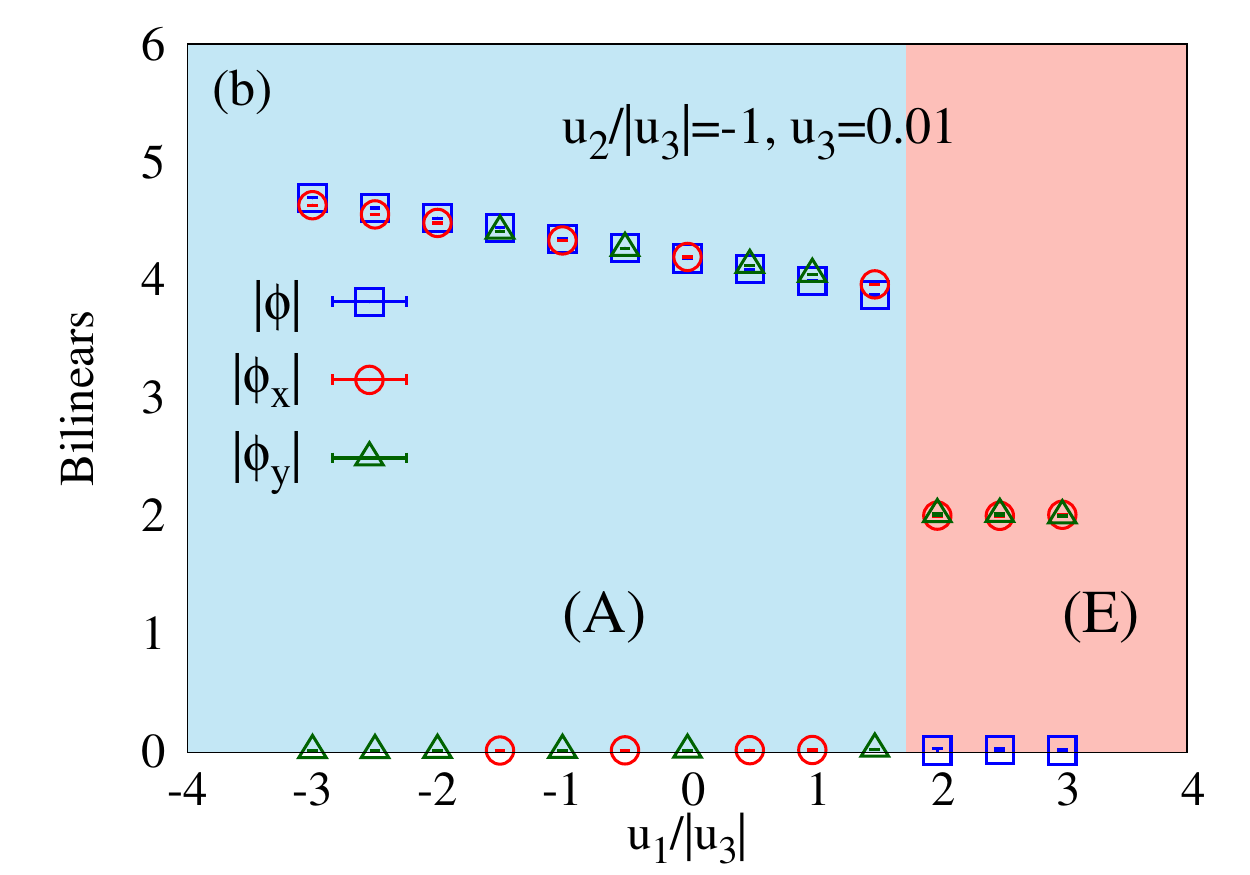}
   \includegraphics[width=0.37\textwidth]{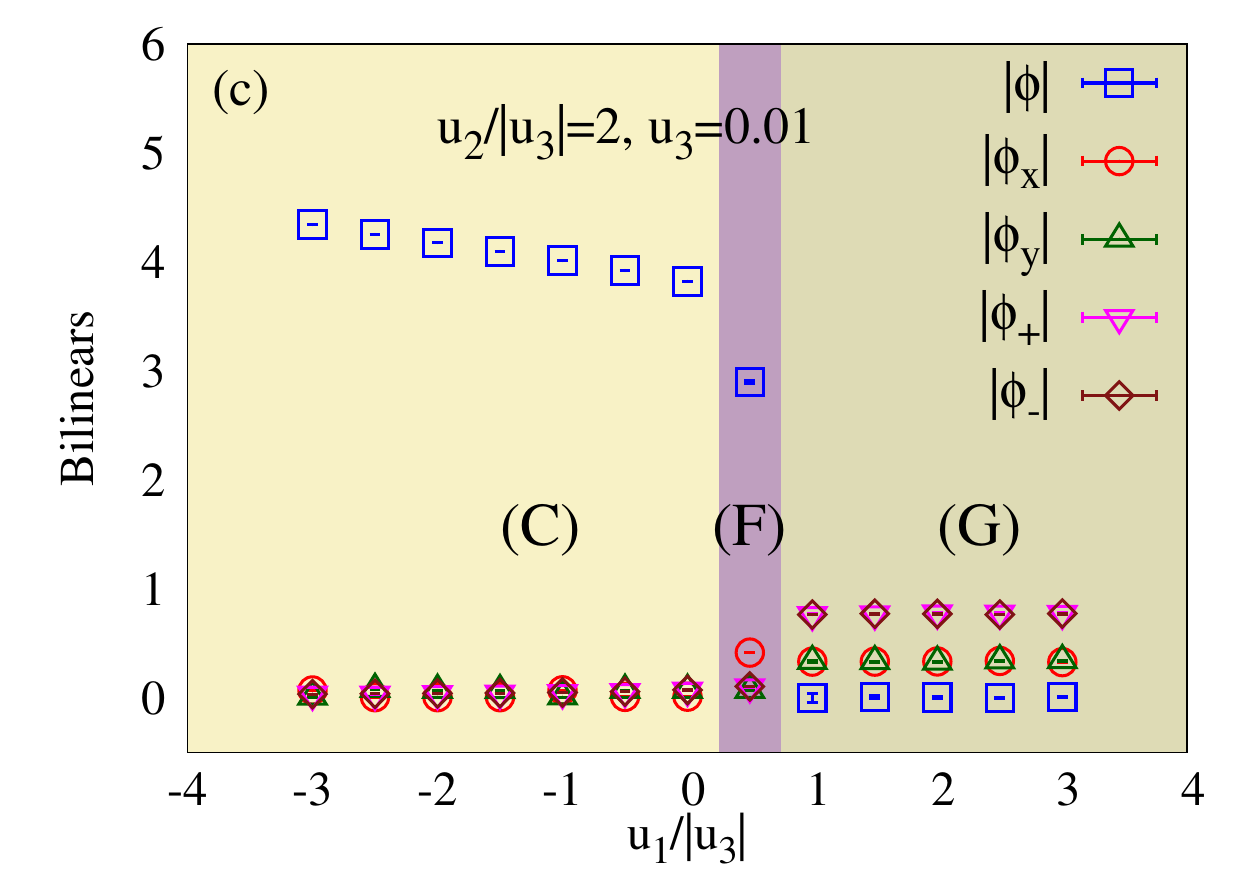}
   \includegraphics[width=0.37\textwidth]{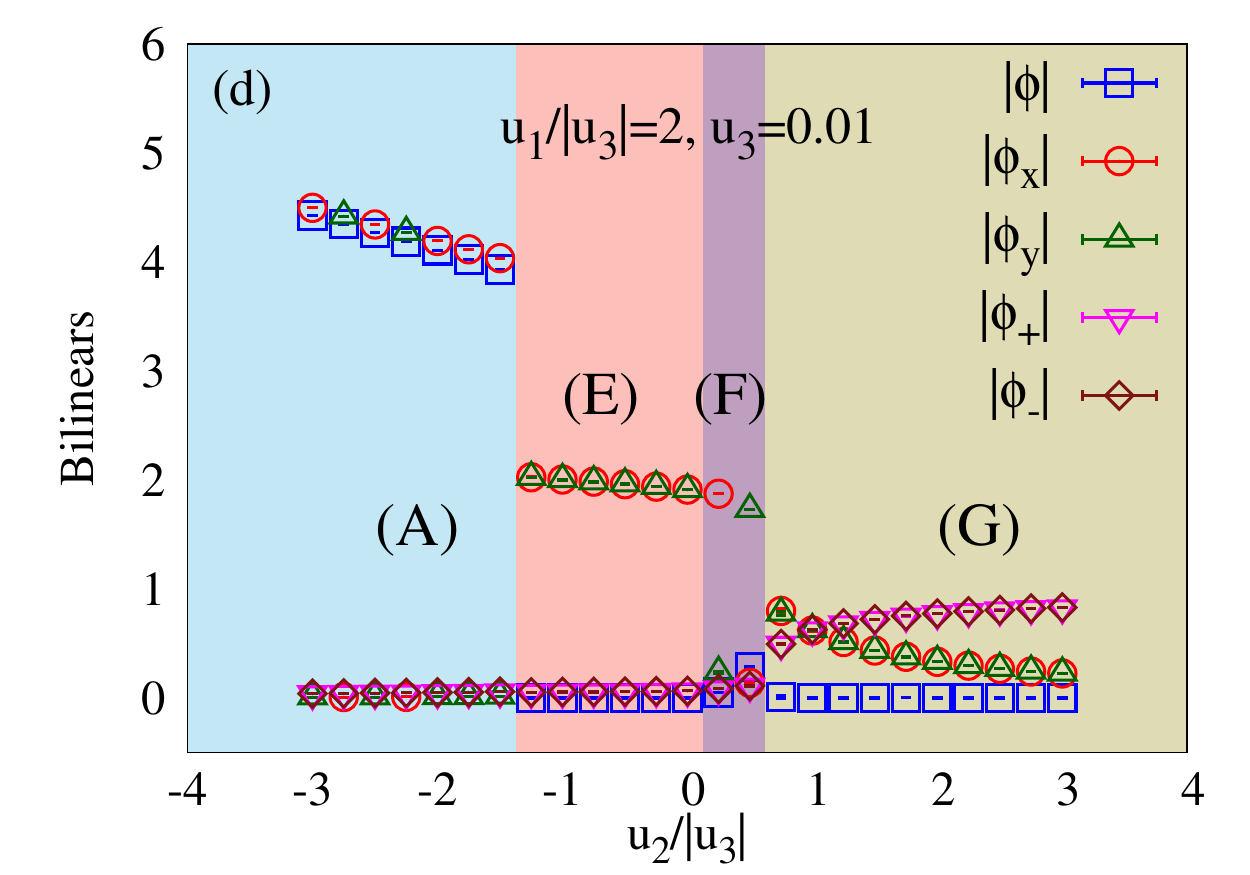}  
  \caption{Expectation values of bilinears across the phase diagram for $u_3>0$. 
	The results have been obtained from simulations on $12^3$ lattices.}\label{fig:pos_u3}
 \end{figure} 
 
\subsection{Simulation details}
We perform Monte Carlo simulations of the gauge theory with adjoint Higgs fields. 
We have used Hybrid Monte Carlo (HMC) algorithm for generating configurations of the 
gauge and Higgs fields and have measured the expectation values of various 
bilinears in Eq.~\ref{Eq:bilinear}. Our preliminary results include measurements 
on $8^3$ and $12^3$ lattices. 
The gauge coupling is set to $\beta=8.0$ for all simulations. In order 
to be in the stable broken phase, the lattice couplings $\kappa$ and $\lambda$ 
(related to the continuum couplings $s$ and $u_0$ by Eq.~\ref{Eq:cont_lat_relations_nh4}) 
are chosen to be $3.0$ and $1.0$ 
respectively. To study the predicted phase diagrams in Fig.~\ref{fig:meanfld}, we 
set the lattice quartic coupling $\hat{u_3}$ to be $0.09$ and $-0.09$, 
and vary $\hat{u}_{1,2}$ such that the ratios 
$\hat{u_1}/|\hat{u_3}|$ and $\hat{u_2}/|\hat{u_3}|$ are in the range $[-3.5:3.5]$.
 
 \begin{figure}
  \centering
   \includegraphics[width=0.37\textwidth]{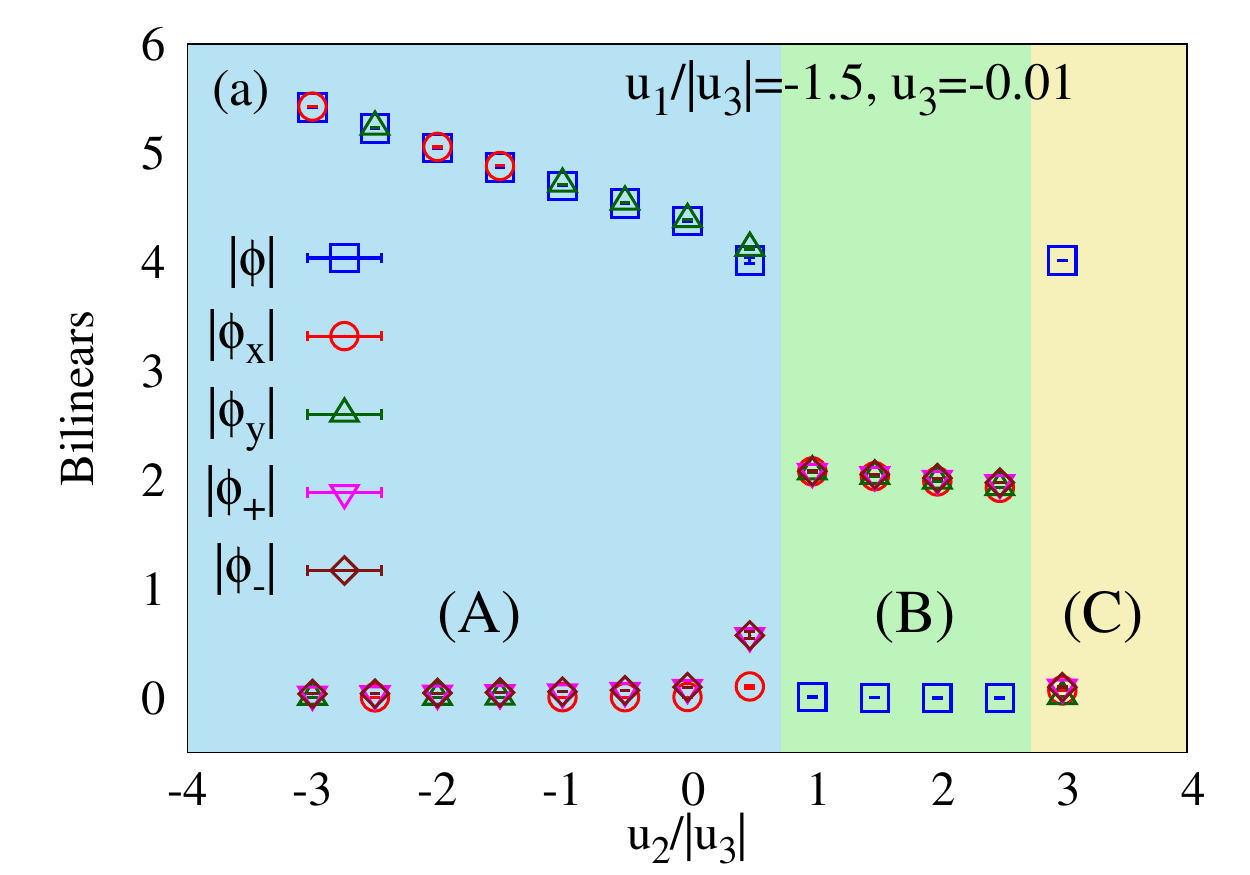}
   \includegraphics[width=0.37\textwidth]{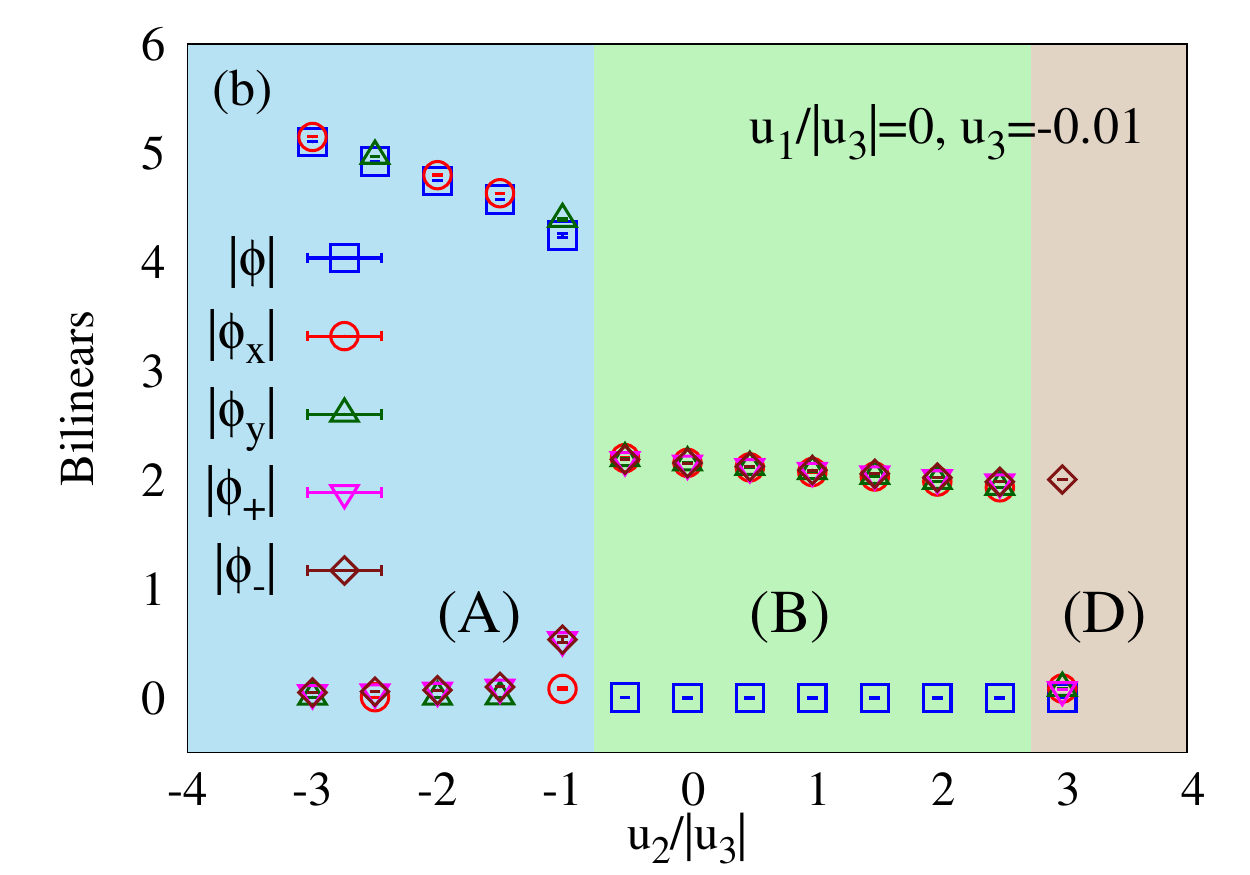}
   \includegraphics[width=0.37\textwidth]{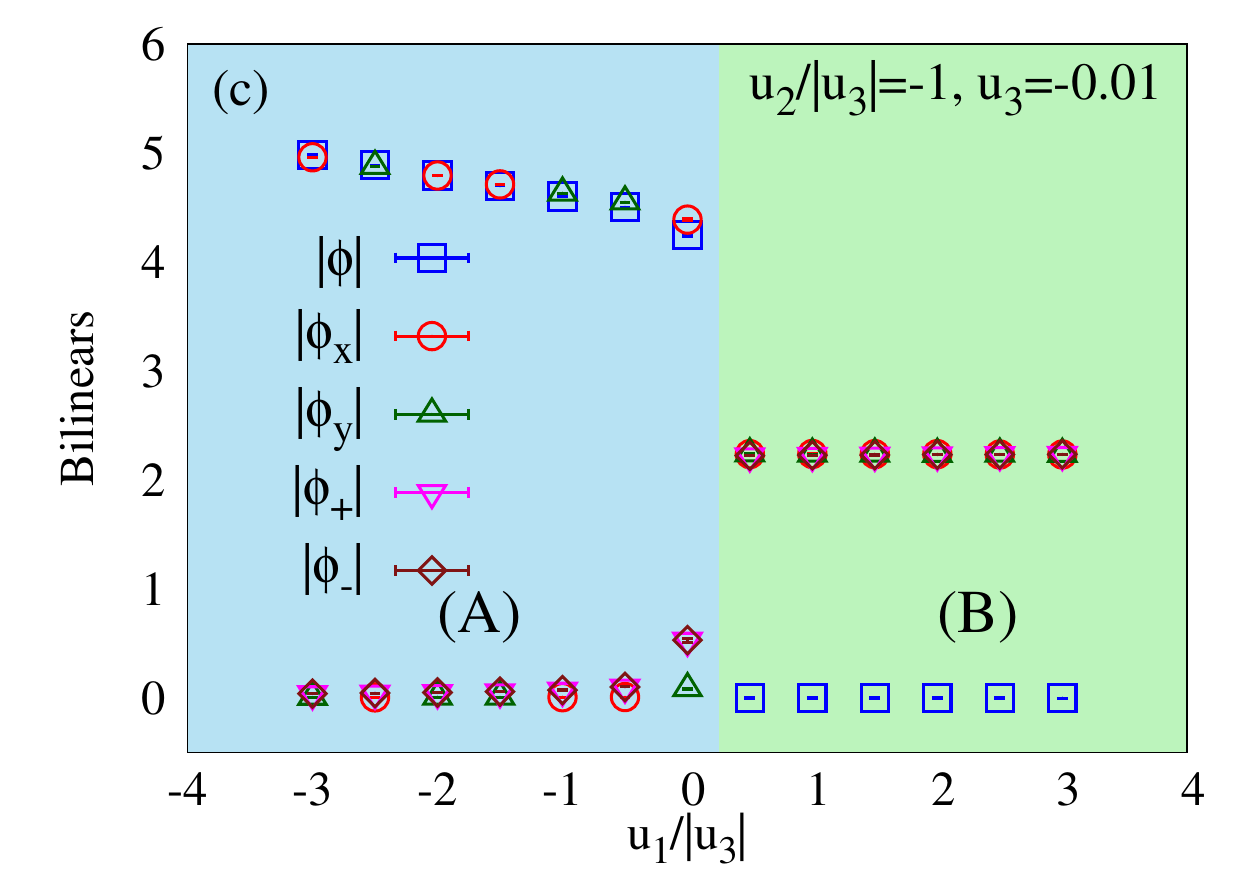}
   \includegraphics[width=0.37\textwidth]{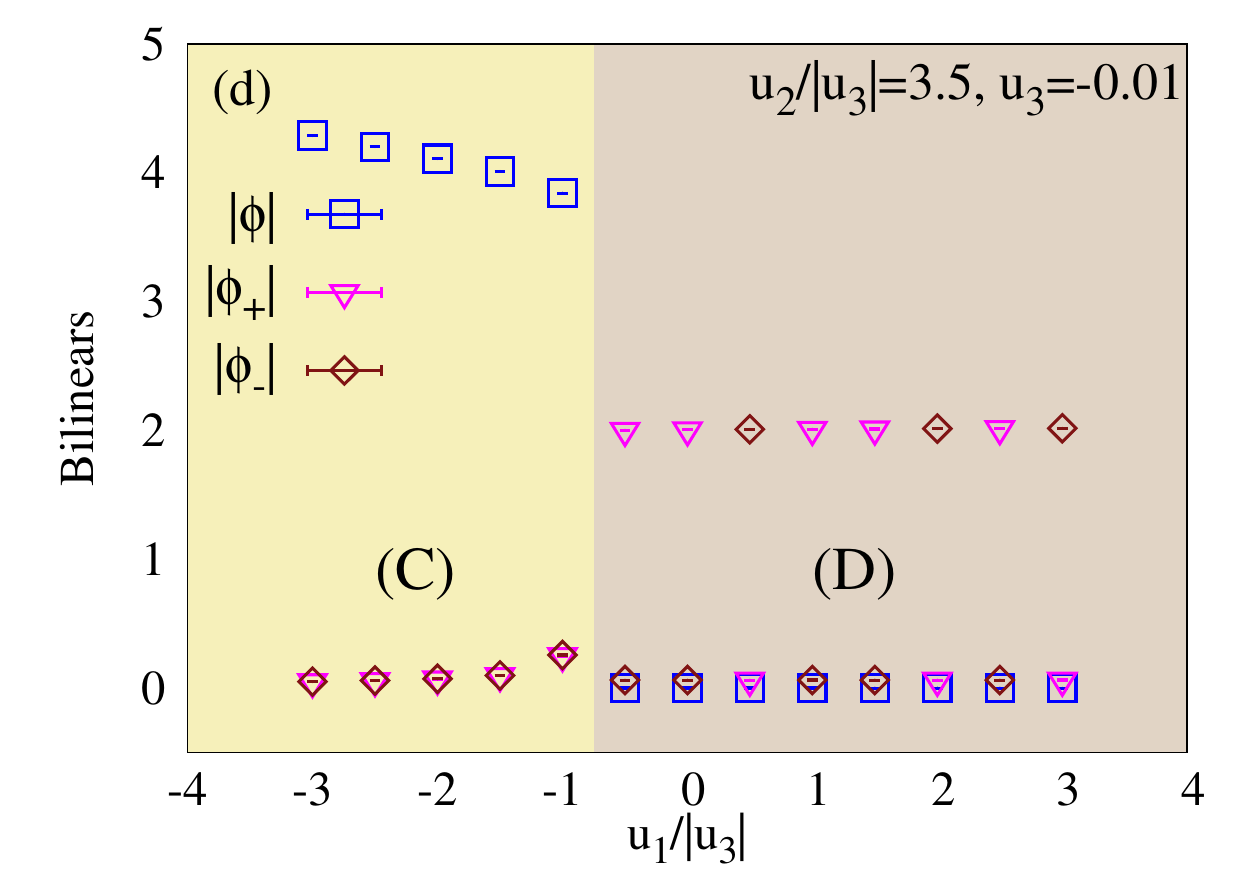}  
  \caption{Results for expectation values of bilinears across the phase diagram for $u_3<0$ obtained 
	 from simulations on $12^3$ lattices.}\label{fig:neg_u3}
 \end{figure} 
 
\subsection{Results}
Our preliminary lattice simulations of the theory appear to qualitatively support 
the mean-field phase diagram. We have performed horizontal and vertical scans of the 
positive/negative $u_3$ phase diagram in Fig.~\ref{fig:meanfld}. 
We plot the variation of the expectation values 
of the bilinears as a function of $u_1/|u_3|$ ($u_2/|u_3|$) 
at fixed $u_2/|u_3|$ ($u_1/|u_3|$). 
 
For $u_3>0$, Fig.~\ref{fig:pos_u3} shows the scanning results. 
In Fig.~\ref{fig:pos_u3} (a), the variation of the 
bilinears\footnote{When we mention bilinears in the following discussion, 
we always mean the expectation values of them.} 
$|\phi|$, $|\Phi_x|$ and $|\Phi_y|$ is shown as a function of $u_2/|u_3|$ 
where the value of $u_1/|u_3|$ is fixed at -1. In agreement with the 
mean-field scenario in Fig~\ref{fig:meanfld} (a), we find a transition between 
phases (A) and (C) around $u_2/|u_3|=0$. The bilinear $|\phi|$ is non-zero in 
both the phases while $|\Phi_x|$ or $|\Phi_y|$, either of which is non-zero in 
(A) vanishes sharply at the transition to phase (C). Next, Fig.~\ref{fig:pos_u3} (b)
depicts the transition between phases (A) and (E) as a function of $u_1/|u_3|$ at a
fixed $u_2/|u_3|=-1$. The bilinear $|\phi|$ is non-zero in phase (A) and goes 
to zero in (E), while the bilinears $|\Phi_x|$ and $|\Phi_y|$ are equal and 
non-zero in phase (E). Fig.~\ref{fig:pos_u3} (c) cuts across the phase diagram (in 
Fig.~\ref{fig:meanfld} (a)) horizontally and shows the transitions between phases 
(C) and (F) followed quickly by the transition to (G) from (F). Along with $|\phi|$, 
$|\Phi_x|$ and $|\Phi_y|$, we plot the bilinears $|\Phi_+|$ and $|\Phi_-|$ 
which are zero in phases (A), (C) and (E). In the transition between 
phases (F) and (G), the bilinear $|\phi|$ goes from non-zero in (F) to zero in (G) 
while $|\Phi_x|$ and $|\Phi_y|$ show similar behavior as in the transition from 
phase (A) to (E). The bilinears $|\Phi_+|$ and $|\Phi_-|$ become equal and non-zero in 
phase (G). Finally, in Fig.~\ref{fig:pos_u3} (d), the variation of the 
bilinears along the vertical direction of the phase diagram 
(cf.~Fig.~\ref{fig:meanfld} (a)) at a fixed $u_1/|u_3|=2$, shows the transitions from 
phase (A)-(E), (E)-(F) and (F)-(G).
 
Next we discuss our results for the case of $u_3<0$ corresponding to the 
mean-field phase diagram in Fig.~\ref{fig:meanfld} (b). The variation of the 
bilinears as a function of $u_2/|u_3|$ at a fixed $u_1/|u_3|=-1.5$ is shown 
in Fig.~\ref{fig:neg_u3} (a). It depicts the transition from phase (A)-(B) and then 
from phase (B)-(C). The bilinear $|\phi|$ vanishes in phase (B) while $|\Phi_x|$, 
$|\Phi_y|$, $|\Phi_+|$ and $|\Phi_-|$ become non-zero and equal\footnote{{The small 
non-zero value of some observables in phase (A)
near the transition from phase (A) - (B) seems to be a
finite volume effect when compared to our ongoing study on
larger volumes.}}.The behavior of 
the observables in phases (A) and (C) are same as in the case of $u_3>0$ discussed earlier.
Along the same direction, Fig.~\ref{fig:neg_u3} (b) displays the transitions from 
phase (A)-(B) and phase (B)-(D) at a fixed value of $u_1/|u_3|=0$. All the bilinears 
vanish in phase (D) except for $|\Phi_+|$ or $|\Phi_-|$. Next in 
Fig.~\ref{fig:neg_u3} (c), the bilinears exhibit a transition from phase (A) to (B) 
as a function of $u_1/|u_3|$ 
at fixed $u_2/|u_3|=-1$. Lastly, the transition from phase (C) to (D) is seen by 
varying the bilinears $|\phi|$, $|\Phi_+|$ and $|\Phi_-|$ with respect to $u_1/|u_3|$ 
at $u_2/|u_3|=3.5$.
 
\section{Two Higgs Doublet Model on the lattice}
\subsection{Background and Action}
{The scalar field structure of the Standard Model (SM) is the simplest 
way to realize electroweak symmetry breaking.}
It consists of a single SU(2) Higgs doublet in the fundamental representation.
On the other hand, the fermion structure of the theory includes 
various families, with mixing involved.
One of the simplest extensions of the SM is the addition of extra 
Higgs fields in the fundamental representation.
In this work we consider the Two Higgs Doublet Model (2HDM). 
{The introduction of a second Higgs doublet is motivated by the
existence of multiple families of fermions in the SM.   It contains
appealing features for model building, such as new sources of
CP violation and a first-order electroweak phase transition (EWPT)  that
drives baryogenesis~\cite{Trodden:1998qg,Fromme:2006cm}.    However,
since no deviation from the SM has been observed at the LHC, 
it is crucial that one of the two Higgs doublets in
2HDM mimics that of the SM.   This is known as the ``alignment''
requirement.  In supersymmetry-inspired 2HDM, it was shown
that the masses of the scalar particles in the second Higgs
doublet are around 5 TeV~\cite{GAMBIT:2017zdo}.  This places the extra
Higgs doublet in the decoupling limit, leading to the fact that it
cannot be discovered in direct searches at the LHC.}

{It has been suggested that the alignment requirement can be fulfilled
without decoupling~\cite{Gunion:2002zf, Carena:2013ooa,
  BhupalDev:2014bir, Hou:2017hiw}.   In this scenario, the masses of
the extra scalar particles are typically at around or below 1 TeV.
Such a realisation of alignment is facilitated with choices of coupling and mixing
strengths in 2HDM.  Notably, as pointed out in
Ref.~\cite{Hou:2017hiw}, it may require certain couplings amongst the
scalar fields to be strong.  Combining this with the trivality
property of these couplings, it is necessary to investigate the
viability of this scenario by examining carefully the relationship
amongst the spectrum, the cut-off scale, and the couplings.}
{To study the above alignment phenomenon without decoupling, as well as
the nature of EWPT, it is necessary to employ
non-perturbative methods to survey the model.  For this purpose,
lattice field theory is the most reliable tool, and is what we pursue
in this work.}

In the lattice action in Eq.~\ref{Eq:genaction}, we set $D=4$ and $N_h=2$. 
In this case, the complex Higgs doublet on the lattice  
$\Phi_n(x) = (\varphi_n^2(x)+i\varphi_n^1(x), \varphi_n^0(x)-i\varphi_n^3(x))$ ($n=1,2$)
is related to the dimensionful continuum field $\phi_n(x)$ as
$\Phi_{n}(x)=a/\sqrt{\kappa}\phi_n(x)$.
For convenience, we combine the four degrees of freedom of the complex doublet 
in the quaternion representation as a matrix,
$\hat{\Phi}_{n}(x) = \frac{1}{\sqrt 2}\sum_{\alpha=0}^{3}\theta_{\alpha}\varphi^{\alpha}_{n}(x)$ 
where
$\theta_{0}=1_{2\times2}$, $\theta_{i}=\sigma_{i}$.
The gauge-Higgs interaction term in Eq.~\ref{Eq:genaction} now contains the 
fundamental $SU(2)$ gauge link $U_\mu(x)$ and is written compactly as 
$-2 \kappa_{n}\Tr \left( \hat{\Phi}_{n}^{\dagger}(x)U_{\mu}(x)\hat{\Phi}_{n}(x+\hat\mu) \right)$.
With the quartic couplings taken to be real, the local Higgs potential takes the form
\begin{align}
  V(\{\hat{\Phi}_n(x)\}) &= 2\mu^{2}\Tr \left( \hat{\Phi}_{1}^{\dagger}(x)\hat{\Phi}_{2}(x) \right) + \xi_{1}\Tr \left( \hat{\Phi}_{1}^{\dagger}(x)\hat{\Phi}_{1}(x) \right)\Tr \left( \hat{\Phi}_{2}^{\dagger}(x)\hat{\Phi}_{2}(x) \right) + \xi_{2}\Tr \left( \hat{\Phi}_{1}^{\dagger}(x)\hat{\Phi}_{2}(x) \right)^{2}  \nonumber \\
				   &  + 2  \Tr \left( \hat{\Phi}_{1}^{\dagger}(x)\hat{\Phi}_{2}(x) \right) \left[\xi_{3}\Tr \left( \hat{\Phi}_{1}^{\dagger}(x)\hat{\Phi}_{1}(x) \right) + \xi_{4}\Tr \left( \hat{\Phi}_{2}^{\dagger}(x)\hat{\Phi}_{2}(x) \right) \right].
\end{align}
It should be noted that the quartic couplings $\lambda_i$ in Eq.~\ref{Eq:genaction} 
are denoted by $\eta_i$ below.

\subsection{Observables and Gradient Flow}

To analyse the phase structure of the system we looked at observables 
which can signal a phase transition, although not being exactly order 
parameters. The squared 
Higgs length, $\expval{\rho_{n}^{2}} = 1/V\expval{\sum_{x} \rho_{n}(x)^{2}}$ with 
$\rho_n^{2} = \det\hat\Phi_n(x)$
for each Higgs doublet was considered. It is non-zero in both the phases but 
increases rapidly in the Higgs phase.
Along with the average plaquette $ P = \frac{1}{12V}\sum_{\square}\Tr{U_{\square}} $, 
the following gauge-invariant links were also considered,
\begin{align}
 L_{\hat{\Phi}_n} = \frac{1}{8V}\sum_{x,\mu}\Tr{\hat{\Phi}_n^{\dagger}(x)U_{\mu}(x)\hat{\Phi}_n(x+\hat\mu)}\,, 
 \quad
 L_{\alpha_n} =\frac{1}{8V}\sum_{x,\mu}\Tr{\alpha_n^{\dagger}(x)U_{\mu}(x)\alpha_n(x+\hat\mu)}\,,
\end{align}
where $\alpha_n$ is the ``angular'' part of the quaternion Higgs field, 
$\hat{\Phi}_{n} = \rho_n \alpha_n$.
The quantity $L_{\alpha}$ is particularly useful since it is bounded between -1 and 1 
and can be used to compare the strength of the transition at different parameter 
regions \cite{Wurtz:2009gf}.

Additionally, since it important to know the scale of the theory, 
gradient flow observables were used.
The gauge invariant action density, 
$\expval{E(x,t)} = -1/2\expval{\Tr G_{\mu\nu}(x,t)G_{\mu\nu}(x,t)}$, 
with $G_{\mu\nu}$ being the flowed gauge field strength
($t$ is the flow time), has been studied.
Two discretizations, namely the plaquette and the clover, for the 
gauge field strength are considered.
The idea of this scale setting is that the fields are flowed until 
the action density reaches a specific value $\mathcal{E}_0$ 
-- this has an associated flow time value $t=t_{0}$, and thus a scale 
$\mu = 1/\sqrt{8t_{0}}$ at which this happens.
By doing this we can determine the relative scales for computations 
performed at different bare couplings. 

In our theory, we are interested
in the Higgs phase of the theory and, we want the  
renormalized gauge coupling at the $W$ boson mass to be
$g^2 \equiv 4\pi\alpha_W \simeq 0.5$ \cite{Langguth:1985dr}.
Using the relation between the renormalized coupling and 
the flowed action density \cite{Luscher:2010iy},
we will obtain the physical theory by requiring
\begin{align}
\label{eq:scale_setting}
  \mathcal{E}_0 \equiv t^2 \expval{E(t)}\Big|_{t=t_0} = \frac{9g^2}{128\pi^2}(1 + \mathcal{O}(g^2))\simeq 0.0036 \Big|_{\mu = 1/\sqrt{8t_0} = m_W},
\end{align}
and the SM Higgs mass being equal to its physical value.

\subsection{Simulation details}
The simulations have been performed using HMC 
algorithm on GPU machines \cite{lattGPU} .
This allows the use of large lattices. 
The results from the HMC code was independently 
verified using a Metropolis algorithm code. 
In our current simulations, the phase transition from 
the confinement to the Higgs phase was 
studied as a function of $\kappa_1=\kappa_2$ 
at different $\beta$ values. 
For all runs, the couplings $\mu^2=0.2$, $\eta_1=\eta_2=0.5$ and 
$\xi_{i}=0.1$ $(i=1,\ldots,4)$ have been kept fixed. 

\subsection{Results}
\begin{figure}
\centering 
\includegraphics[width=0.4\textwidth]{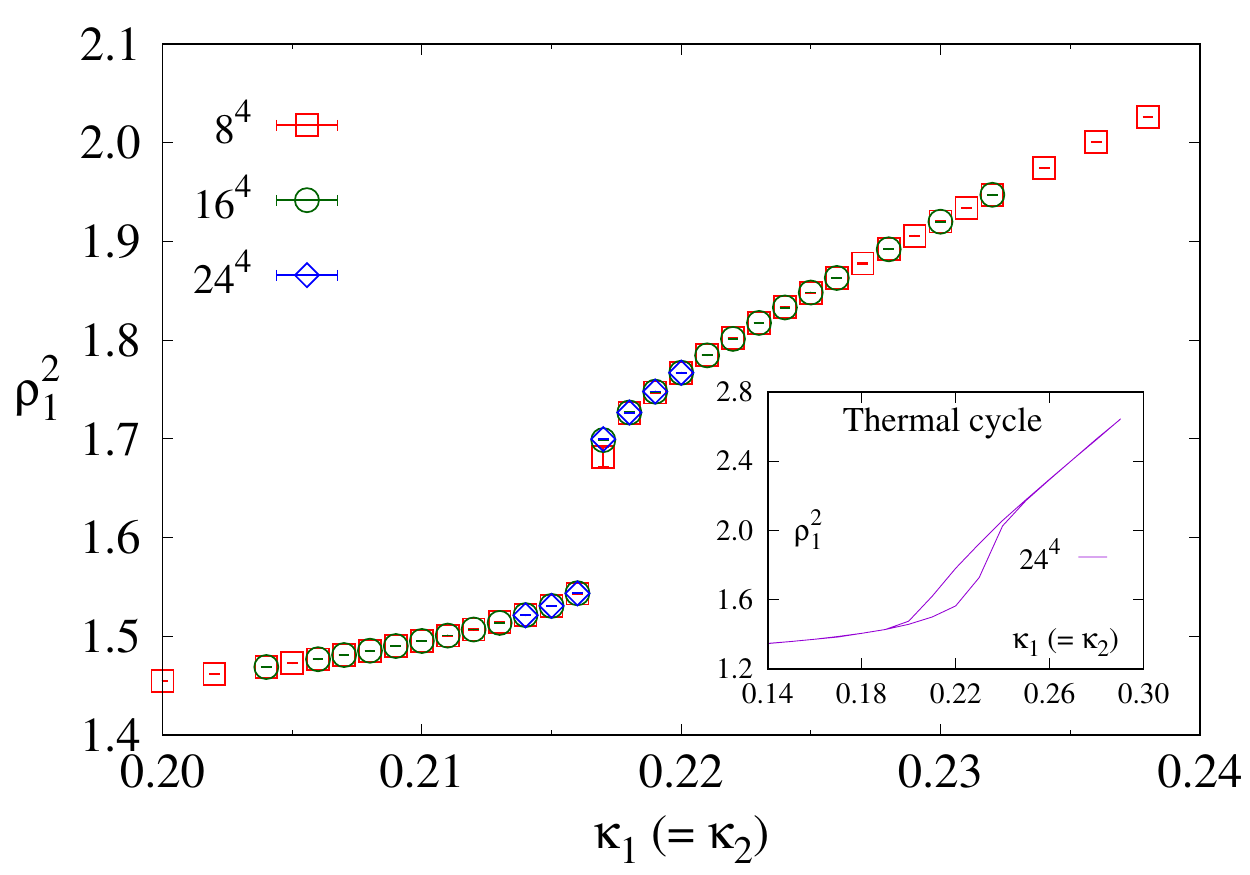}
\includegraphics[width=0.4\textwidth]{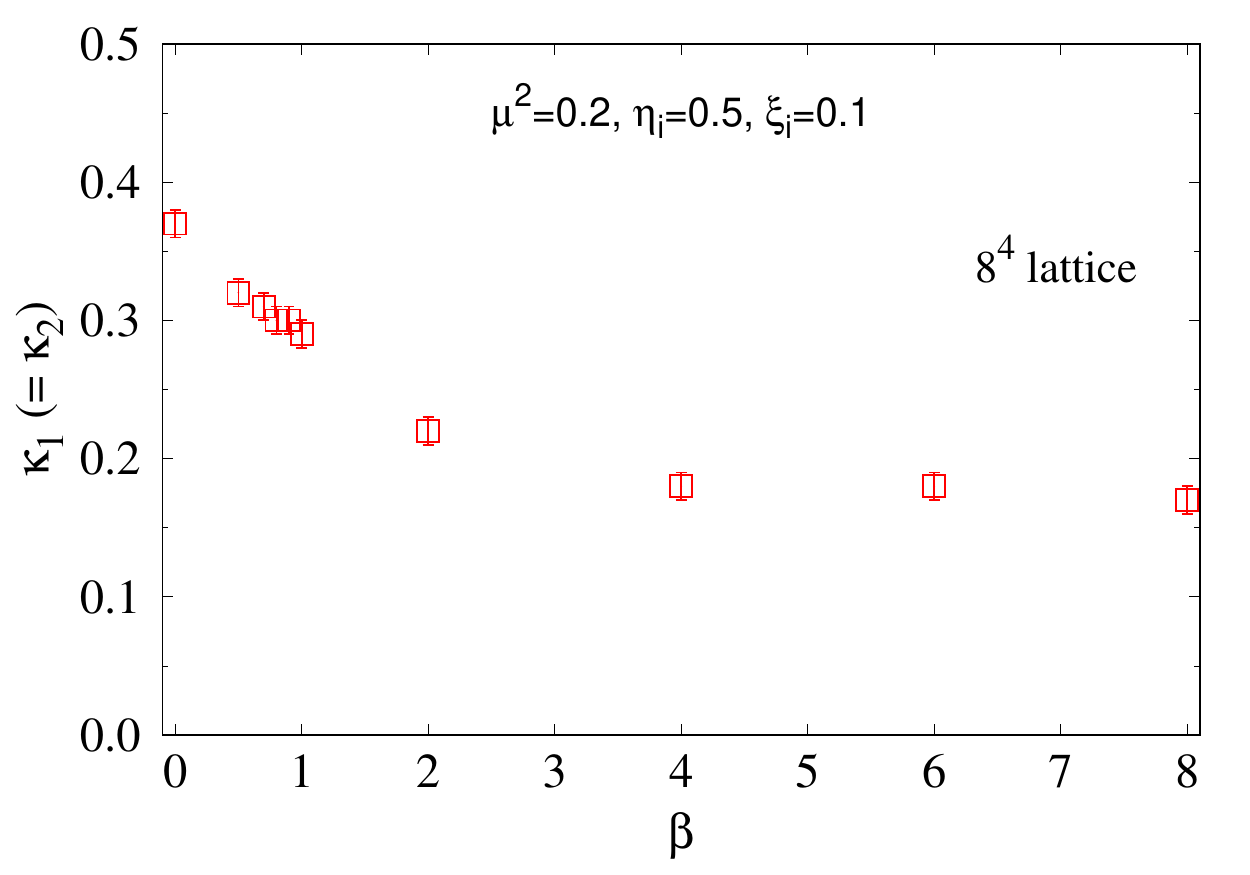} 
\caption{\textit{Left}: Preliminary result of the variation of the Higgs length $\rho_1^2$ with respect to $\kappa_1$ ($=\kappa_2$) at $\beta=2.0$ on $8^4$, $16^4$ 
and $24^4$ lattices. \textit{Inset}: Thermal cycle of $\rho_1^2$ obtained on a $24^4$ lattice displaying a hysteresis behavior.
\textit{Right}: Preliminary results of the phase transition points in the  
$\beta$ - $\kappa_1$ ($=\kappa_2$) plane found on $8^4$ lattices. 
Results in both plots have been obtained at fixed $\mu^2=0.2, \eta_i=0.5$ and $\xi_i=0.1$.} \label{fig:twoHiggs}
\end{figure}
A preliminary analysis of the phase transition, obtained by changing the 
$\kappa_1=\kappa_2$ values  
at a constant $\beta$, allowed us to identify the 
Higgs/confinement phases in the space of bare parameters.
To evaluate the phase transition line in the $\beta$ - $\kappa_1$/$\kappa_2$ plane,
we calculate the gauge invariant observables on $8^4$, $16^4$ and $24^4$ lattices. 
We show a preliminary result of the hopping parameter
$\kappa_1$ $(=\kappa_2)$ dependence the Higgs length $\expval{\rho_1^2}$ at 
$\beta=2.0$ on different lattices in the left plot in
Fig.~\ref{fig:twoHiggs}. 
The observable shows a jump indicative of a first-order phase transition 
around $\kappa_1=\kappa_2=0.217$ at this parameter set. In the inset, we show 
results from a thermal cycle run which clearly indicates a hysteresis loop. 
The thermal cycle has been obtained by taking the system through steps of 
$\kappa_1=\kappa_2=0.01$ with only 5 HMC trajectories at each point, starting 
from $\kappa_1=\kappa_2=0.14$ to $0.29$ and coming back. 
The system was allowed to equilibrate
at the end points. At each point in the up or down paths of the cycle, the 
HMC run uses the last configuration from the previous point as the initial start. 
Further analysis via finite size scaling or observation of a 
two-state signal near the transition needs to be obtained to confirm the 
position and order of the transition.
We show the transition points obtained at different values in  
the $(\beta,\kappa_1=\kappa_2)$ phase diagram in 
the right plot of Fig.~\ref{fig:twoHiggs}. {The transition at intermediate $\beta$ 
show a strong jump in the observables (as seen in  
Fig.~\ref{fig:twoHiggs} left) but become gradually continuous at 
smaller or larger $\beta$.}
These results are similar to the case of SU(2) gauge theory 
with a single Higgs doublet \cite{Wurtz:2009gf}. 
\begin{figure}[]
	\begin{minipage}[t]{.5\textwidth}
		\centering
		\scalebox{0.55}{\includegraphics[scale=0.6]{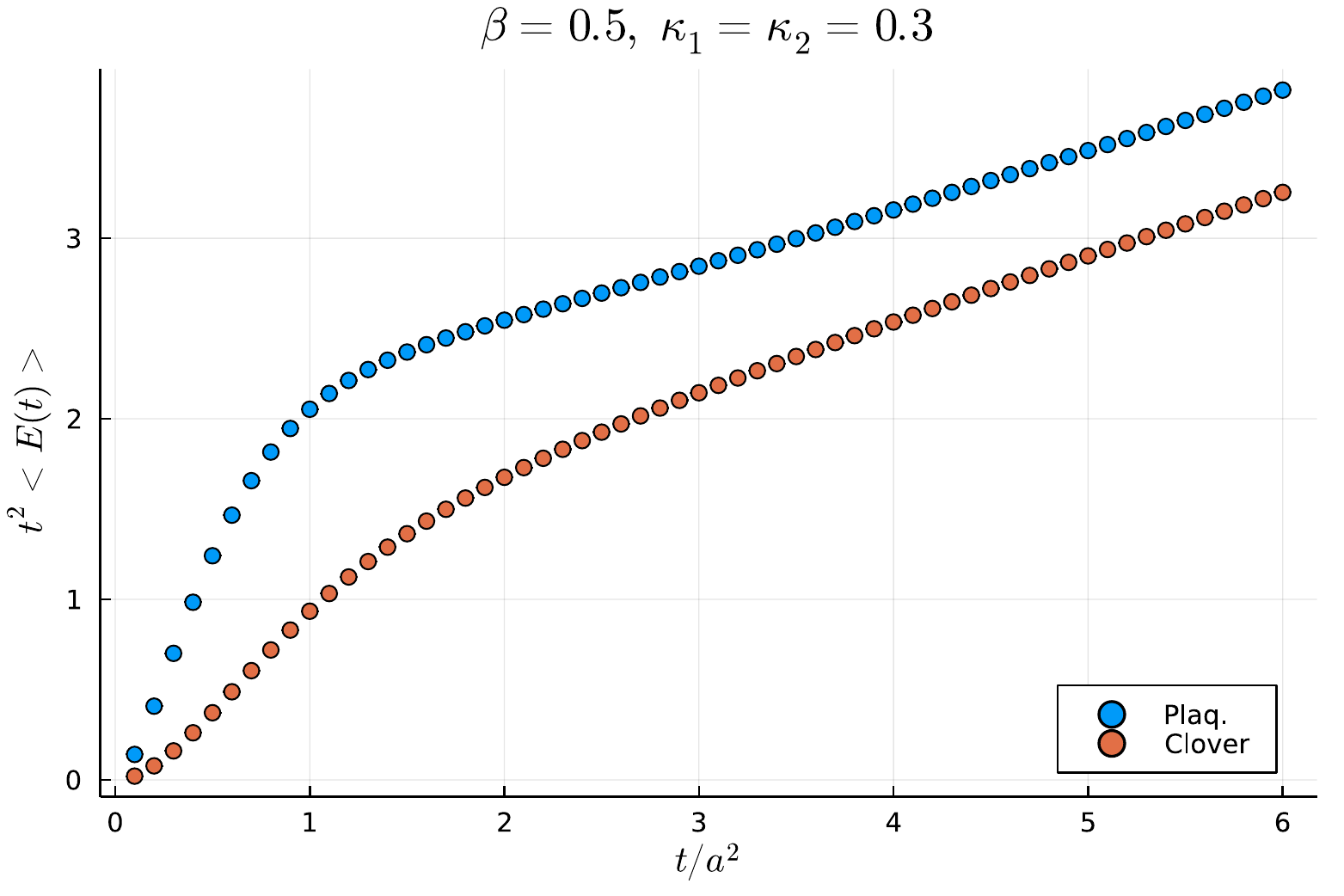}}
	\end{minipage}%
    \begin{minipage}[t]{.5\textwidth}
		\centering
		\scalebox{0.55}{\includegraphics[scale=0.6]{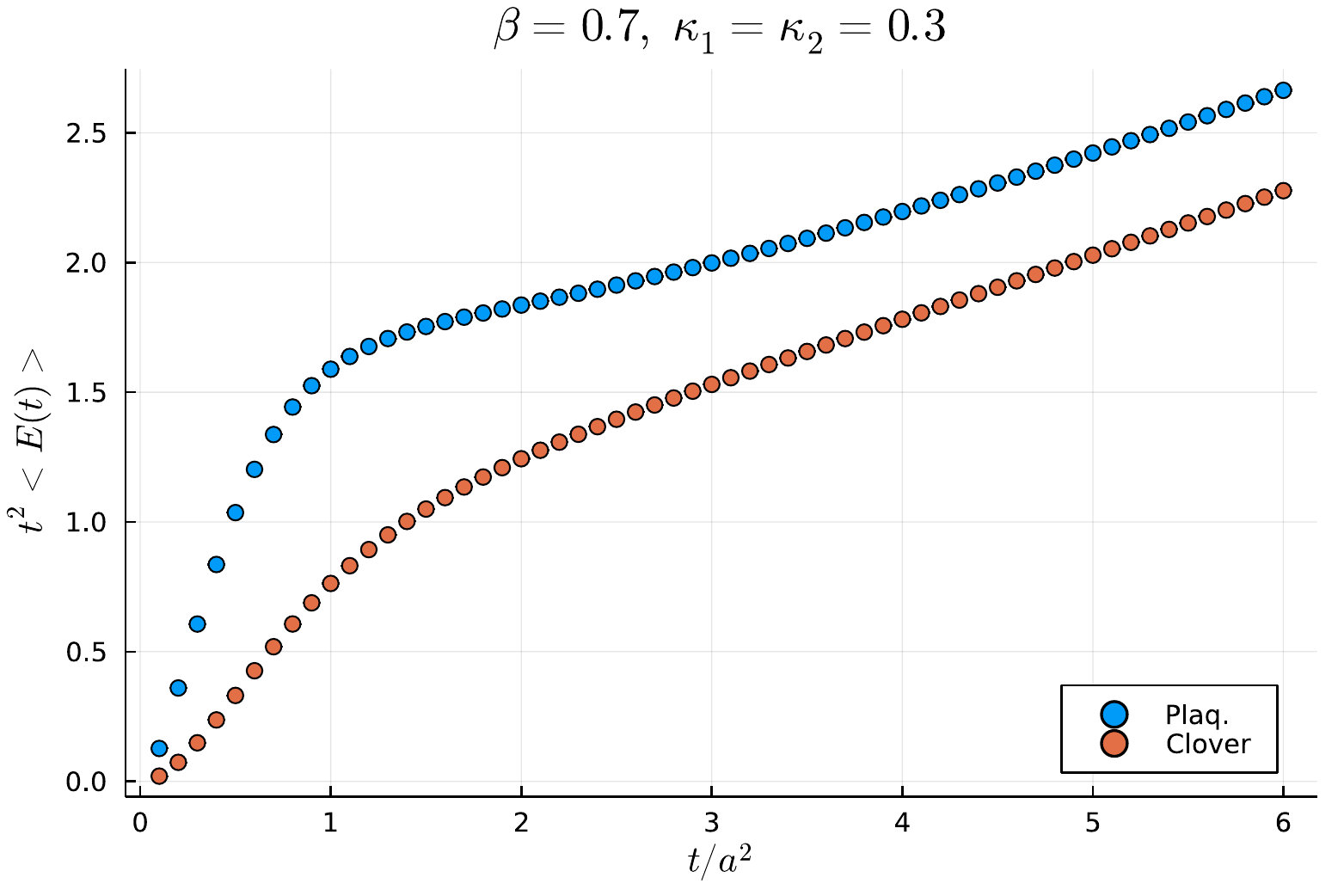}}
      \end{minipage}%
    \\
    \begin{minipage}[t]{.5\textwidth}
		\centering
		\scalebox{0.55}{\includegraphics[scale=0.6]{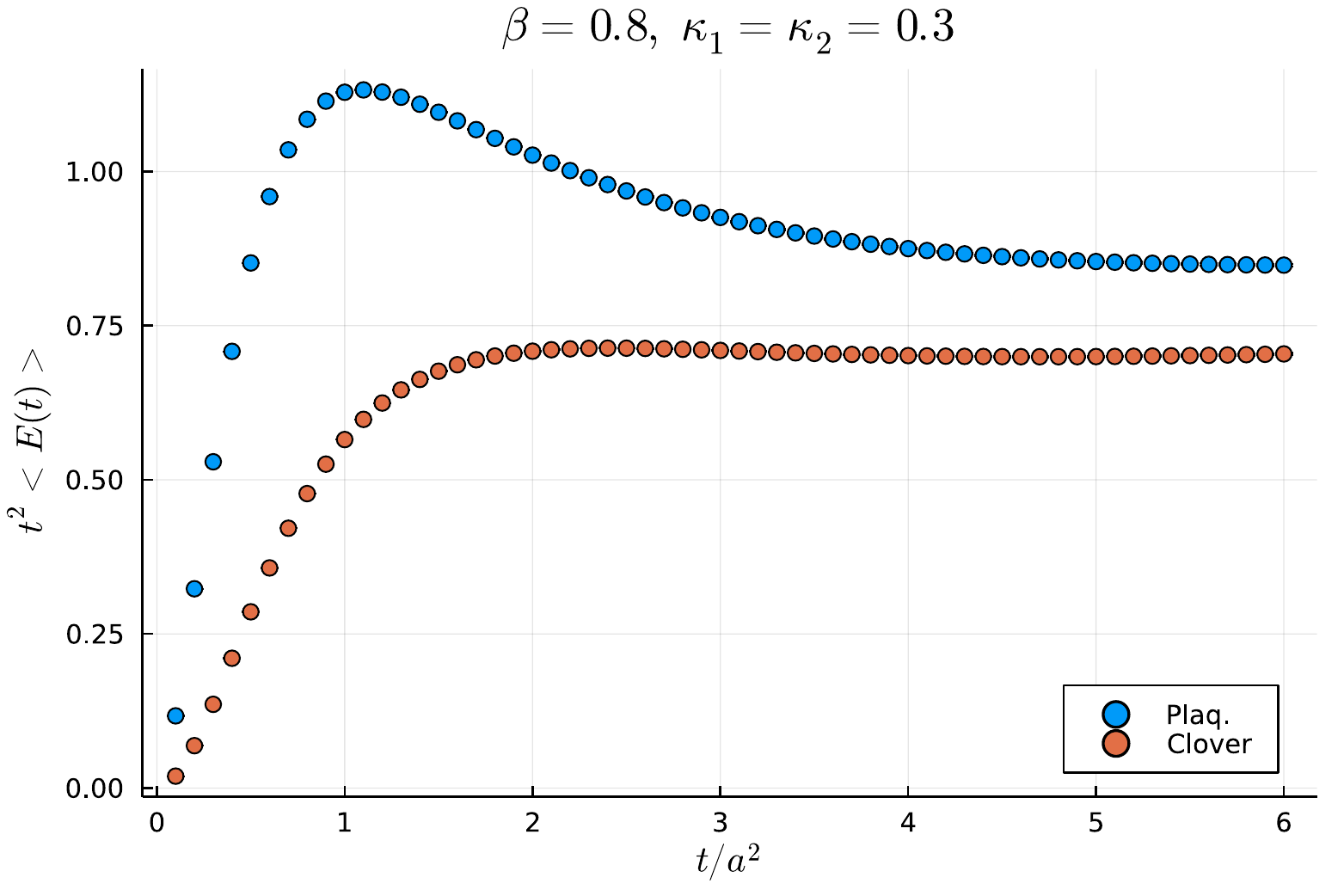}}
	\end{minipage}%
    \begin{minipage}[t]{.5\textwidth}
		\centering
		\scalebox{0.55}{\includegraphics[scale=0.6]{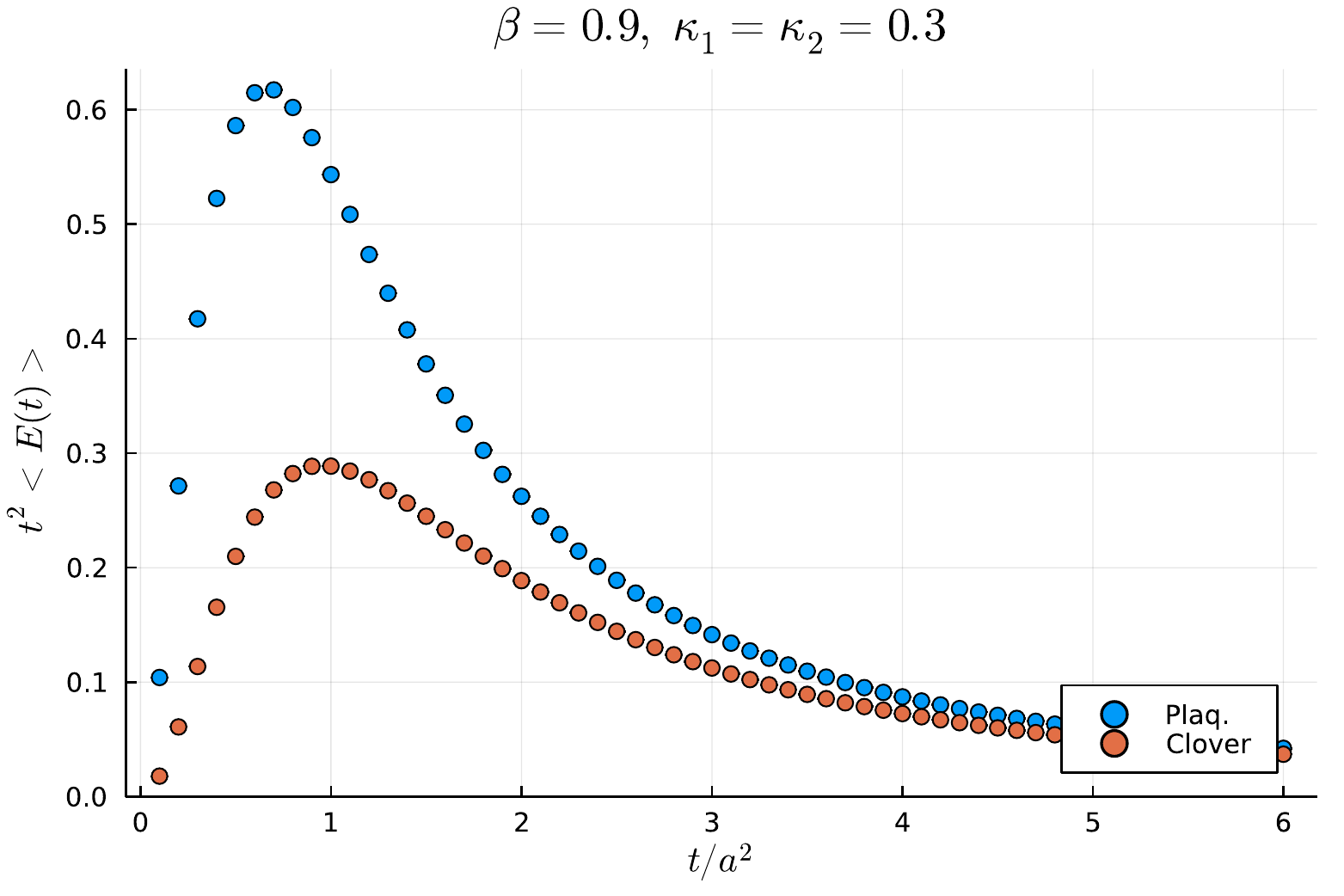}}
	\end{minipage}%
 \caption{$t^{2}\expval{E(t)}$ for $\kappa_{1}=\kappa_2=0.3$ and various $\beta$ values 
	from 500 flow measurements. For all the runs, the quadratic coupling is $\mu^2=0.2$ and 
 the quartic couplings are $\eta_i=0.5$ and $\xi_i=0.1$.}
	\label{fig:t2E for k 0.3}
\end{figure}

{As for scale setting, our preliminary investigations 
of the flow of the plaquette and clover action density, $t^{2}\expval{E(t)}$,
as a function of the flow time, $t/a^2$, are shown in Fig.~\ref{fig:t2E
  for k 0.3}. The quantity $t^{2}\expval{E(t)}$ is
proportional to the gauge coupling in the gradient-flow scheme 
[see Eq.~(\ref{eq:scale_setting})].
Results presented in this figure indicate a significant change in 
behaviour of the gauge coupling with flow time 
as one increases the $\beta$ at a fixed $\kappa_1=\kappa_2$. 
At smaller $\beta=0.5,0.7$, the system is in the confining phase and 
it goes over to the Higgs phase as $\beta$ is increased,
(see Fig.~\ref{fig:twoHiggs}).
Plots in the top row of Fig.~\ref{fig:t2E for k 0.3} exhibit
$t^{2}\expval{E(t)}$ as a function of $t$ in the confining phase, 
In this phase, behaviour similar to that of QCD is observed, although the lattice 
artefacts are sizeable in simulations with these two $\beta$ values.
The bottom-row plots in Fig.~\ref{fig:t2E for k 0.3} demonstrate the
same quantity in the Higgs phase. It is noticed that at long
distance, the gauge coupling in the gradient-flow scheme decreases
with the renormalisation scale. Qualitative understanding of this
behaviour can be achieved by studying the one-loop $\beta{-}$ function
of the gauge coupling in the SU(N) gauge theory coupled to
$n_{s}$ scalar fields,
\begin{equation}
\label{eq:beta_func}
 \beta_{{\mathrm{SU(N)-scalar}}} = \mu \frac{d g}{ d \mu} = -\frac{b_{0} g^{3}}{16
 \pi^{2}} + {\mathcal{O}}(g^{5}) \, , \,\, {\mathrm{where}} \,\, b_{0} =
\frac{11N - n_{s}}{3}\, .
\end{equation}
Below the $M_{W}$ threshold in the Higgs phase, the gauge bosons are
integrated out, and $N=0$ in $b_{0}$, resulting in a positive
$\beta_{{\mathrm{SU(N)-scalar}}}$. The short-distance ($t/a^{2}
\rightarrow 0$) behaviour shown in these plots seems to be consistent
with the presence of asymptotic freedom.  Although this can also be
understood by having $N=2$ in Eq.~(\ref{eq:beta_func}) (above the $M_{W}$
threshold), lattice artefacts are very significant in
this regime, where further detailed investigation  
is needed in the future.
}

\section{Conclusions and outlook}
We study the $3d$ $SU(2)$ gauge theory with four flavors of Higgs fields 
transforming under the adjoint representation. The theory was proposed 
to describe the physics of hole-doped cuprate systems near optimal doping 
in Ref.~\cite{Sachdev:2018nbk}. In our lattice study, we have qualitatively 
confirmed the mean-field predictions of the theory in the Higgs phase. 
We show the existence of the various broken phases signalled by the 
expectation values of gauge-invariant order parameters bilinear in the Higgs fields.
We are currently working on identifying the order of the phase transitions and 
finding out a purported deconfined quantum critical point, as predicted in 
\cite{Sachdev:2018nbk}. 

In the lattice 2HDM,
we calculate gauge invariant observables to investigate phase transition lines 
in the $(\beta,\kappa_1=\kappa_2)$ phase diagram.
This is the first-time study of complete potential with real couplings.
Our preliminary results indicate the existence of first-order transition separating
the confinement and Higgs phases of the theory. However, this needs to be 
established.  
We implement the gradient flow for the $SU(2)$ gauge fields in the 2HDM 
for setting the scale. 
Our goal is to study the mass spectrum of the two Higgs doublet model with the 
physical $W$ boson and Higgs boson masses. 

\section{Acknowledgements}
The work of AH is supported by the Taiwanese
MoST grant 110-2811-M-A49-501-MY2 and 111-2639-M-002-004-ASP.
The work of CJDL is supported by the Taiwanese
MoST grant 109-2112-M-009-006-MY3 and 111-2639-M-002-004-ASP.
AR and GT acknowledge support from the Generalitat Valenciana
(genT program CIDEGENT/2019/040) and the Ministerio de Ciencia e
Innovacion  PID2020-113644GB-I00. 
The work of MS is supported by the Taiwanese
MoST grant NSTC 110-2811-M-002-582-MY2, NSTC 108-2112-M-002-020-MY3 and 111-2124-M-002-013-.
The authors gratefully acknowledge
the computer resources at Artemisa, funded by the European Union ERDF
and Comunitat Valenciana as well as the technical support provided by
the Instituto de Física Corpuscular, IFIC (CSIC-UV), and the 
computer resources at AS, NTHU, NTU, NYCU and DESY, Zeuthen.

\end{document}